# Thermodynamic of photoluminescence far from the radiative limit

A. Manor[1], M. Kurtulik[1], C. Rotschild[1,2]

*1 Russell Berrie Nanotechnology Institute, Technion − Israel Institute of Technology, Haifa 32000, Israel*

*2 Department of Mechanical Engineering, Technion − Israel Institute of Technology, Haifa 32000, Israel*

The radiance of thermal emission, as described by Planck's law, depends only on the emissivity and temperature of a body, and increases monotonically with temperature rise at any emitted wavelength. Non-thermal radiation, such as photoluminescence (PL), is a fundamental light–matter interaction that conventionally involves the absorption of an energetic photon, thermalization, and the emission of a redshifted photon. Until recently, the role of rate conservation when thermal excitation is significant, has not been studied in any non-thermal radiation. A question: What is the overall emission rate if a high quantum efficiency (QE), PL material, is heated to a temperature where it thermally emits a rate of 50 photons/sec at its bend edge, while in parallel is PL excited at a rate of 100 photons/sec. Recently, we discovered that the answer is an overall rate of 100 blueshifted photons/sec [1]. In contrast to thermal emission, the PL rate is conserved with temperature increase, while each photon is blueshifted. Further rise in temperature leads to an abrupt transition to thermal emission where the photon rate increases sharply. We also demonstrated how endothermic-PL generates orders of magnitude more energetic photons than thermal emission at similar temperatures. These findings show that PL is an ideal optical heat pump, and can harvests thermal losses in photovoltaics with theoretical maximal efficiency of 70% [2], and practical device that aims to reach 48% efficiency [3]. Here we study, for the first time, the emission rate of PL radiation at moderate QE where non-radiative processes dominate the dynamics. Though conservation of photon rate doesn't apply, we predict that the emission rate at 100% QE is an upper limit for the overall emission rate regardless of the QE. Also, the transition temperature to thermal emission is QE independent. These fundamentals of light and heat are expected to have high impact on light and energy science.

## Introduction

The fundamental physics that governs the interplay between PL and thermal emissions is expressed by the generalized Planck law, describing non-thermal emission [1,4]:

$$R(\hbar\omega, T, \mu) = \varepsilon(\hbar\omega) \cdot \frac{(\hbar\omega)^2}{4\pi^2\hbar^3 c^2} \frac{1}{e^{\frac{\hbar\omega-\mu}{K_B T}}-1} \cong R_0 \cdot e^{\frac{\mu}{K_B T}} \quad (1)$$

Where $R$ is the emitted photon flux (photons per second per unit area). Here, $T$ is the temperature, $\varepsilon$ is the emissivity, $\hbar\omega$ is the photon energy, $K_b$ is Boltzmann's constant and $\mu$ is the chemical potential. The corresponding emitted energy rate is defined by $E(\hbar\omega, T, \mu) = R(\hbar\omega, T, \mu) \cdot \hbar\omega$. The chemical potential $\mu > 0$ defines the level of excitation above the system's thermal equilibrium, $R_0$, and is frequency-invariant at the spectral band wherein thermalization equalizes excitation levels between modes. This is true for excited electrons in the conduction band of solid-state semiconductors as well as for excited electrons in isolated molecules[1]. For semiconductors, $\mu$ is the gap between the quasi-fermi-levels that is opened upon excitation. By its definition, for a fixed excitation rate, as temperature increases, $\mu$ is reduced and when $\mu = 0$ the radiation is reduced to thermal emission, $R_0$. Thermodynamically, the chemical potential is defined as long as the number of particles is conserved, which for PL means constant quantum efficiency (QE), i.e. the ratio between the emitted and absorbed photon rates. Eq. 1 describes the excitation of electrons at a specific band where $\mu$ is constant. Although this theoretical formalism and the thermodynamics of optical refrigeration [5],[6] are well established, the regime where thermal excitation is comparable to the level of photonic excitation was never explored. We start by noticing that any additional thermal excitation of electrons from the ground state, i.e. thermal emission that rapidly grows with the rise of temperature, cannot be added to the PL rate described by Eq. 1. This is because such a sum of emissions would result in total thermal emission (at $\mu = 0$) that exceeds the Black Body radiation. In another intuitive description; the expectation that a low radiance thermal source (heating below critical temperature) increases high radiance PL is similar to the expectation that a cold body heats a hot body. This is of course a violation of the 2nd law of thermodynamics.

With this in mind, we start with simulating the PL evolution of an ideal material, under constant optical pump rate and temperature increase. Physically, as will be shown in the experimental section, this corresponds to a situation where the material is optically pumped at its absorption resonance, generating constant PL, with additional variable heat flux that controls the temperature. For the sake of generality the material is chosen to have a band-like emissivity function, as shown in figure 1a (inset). This emissivity function can describe both materials with discrete energy gaps, such as small molecules, and semiconductors (by expending the emissivity into the high energy spectrum). As an example, the emissivity function is chosen to be unity between 1.3eV and 1.7eV and zero elsewhere. In addition, at this stage, the PL is assumed to have unity quantum efficiency (QE) and only radiative heat transfer is accounted for. We solve Eq. 1 by balancing the incoming and outgoing photonic and energy rates, at steady state. For a given incoming photon and energy rate the solution uniquely defines the thermodynamic state of the PL absorber, which is characterized by its quantities $T$ and $\mu$. The only way to conserve both the PL and energy rates is if each emitted photon is blue-shifted with the increase in pumped heat. Figure 1a presents the evolution of emission spectrum and chemical potential (inset) as function of temperature. Figure 1b presents the total emitted photon rate (inset) and the rate of photons with energy above 1.45eV in the case of endothermic PL (Blue line) and thermal emission (Red line). The thermal emission is calculated by setting $\mu = 0$, and applying only the energy balance. As evident, at low temperatures, the emission's line shape at the band-edge is narrow, and is blue-shifted with temperature increase (Fig. 1a), while the total emitted photon rate is conserved (Fig. 1b inset). Remarkably, in opposite to thermal emission, this process is characterized by the **reduction of photon rate near the band edge**, where electrons are being thermally-pumped to the high energy regime as long as $\mu > 0$. The red portion of the emission in figure 1a represents the thermal population, $R_0$. At low temperatures the PL photon rate is far above the rate of thermal emission, while $R_0$ increases and becomes significant at high temperatures. The temperature rise leads to the reduction in the chemical potential, according to the relation:

$$\mu(T) = K_B T \cdot \ln\left(\frac{\int R \cdot d(\hbar\omega)}{\int R_0 \cdot d(\hbar\omega)}\right) \qquad (2)$$

This trend continues until $\mu = 0$, where the emission becomes purely thermal. For the computation in this case the constraint for balance between the absorbed and PL photon rates is removed. Further rise in temperature results in a sharp increase of the photon rate at all wavelengths. Examining the generation rate of photons with energy above 1.45 eV, corresponding to λ<850nm (figure 1b) shows the emitted rate of energetic photons in the endothermic PL case (Blue line) is orders of magnitude greater than in thermal emission under the same temperatures (Red line). At µ = 0, both energetic photon rates converge.

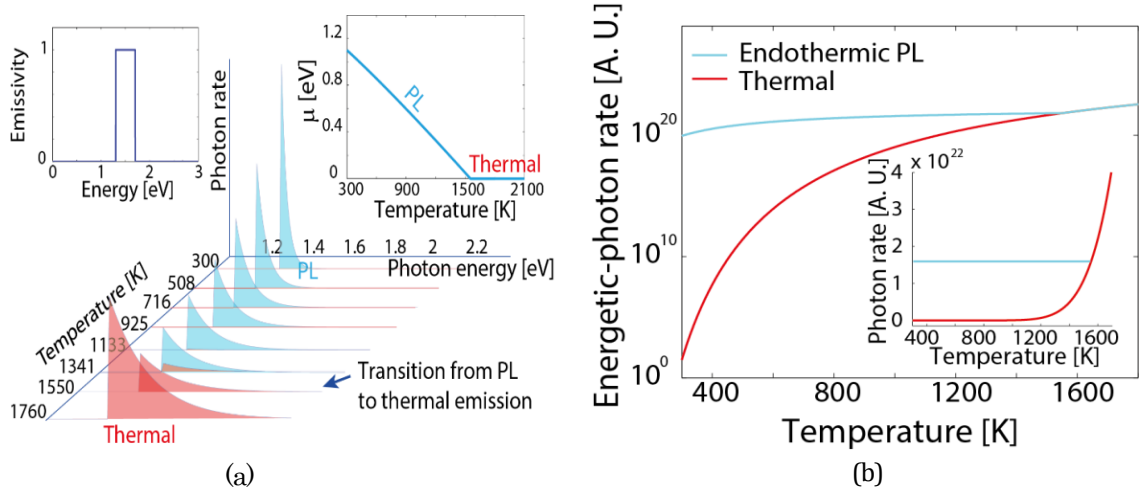

Fig. 1. (a) Emission evolution of PL material with temperature. Insets: The emissivity function and the chemical potential temperature dependence (b) Emission rates of energetic photons and total photons rate (inset) for PL (Blue line) and thermal emission (Red line) at various temperatures.

Moving to the experimental demonstration, the need for high QE PL at high temperatures limits the use of solid-state semiconductors, due to reduction in their QE by temperature-dependent non-radiative recombination mechanisms[7]. On the other hand, rare-earth ions such as Neodymium and Ytterbium are an excellent choice of materials as their electrons are localized and insulated from interactions. This results in the conservation of their high QE at extremely high temperatures[8]. Due to the lack of interaction between electrons, each energy gap can be populated at different $\mu$ values, but thermalization equalizes $\mu$ within each spectral band. We experimentally study the transition between the PL and thermal emission at the 905 nm fluorescence band of Neodymium ($Nd^{+3}$) doped silica fiber tip under 532 nm PL excitation of few mW. The 905nm emission corresponds to the transition between the $4F_{3/2}$ and $4I_{9/2}$ energy levels of the $Nd^{3+}$ system, the latter being the ground state, which is essential for maximizing the thermal radiation signal. In order to control the heat load separately from the PL excitation, we use a $CO_2$ laser operating at 10.6µm wavelength with power levels up to 150 mW. At this wavelength the photons are efficiently absorbed by the silica matrix[9], and converted to a constant heat flow. Although the 532nm pump is exothermic in the sense that the absorbed energy is higher than the emitted PL energy, the heat associated with the few mW pump thermalization is negligible, in comparison with the heat load imposed by the CO2 laser. Therefore, the PL harvests the thermal energy pumped by the CO2 laser, thus it is endothermic. The temperature and the chemical potential are uniquely defined by these two pumps. The experimental setup is sketched in figure 2a. The power spectrum is measured by a calibrated spectrometer. The weak PL excitation at 532nm is kept constant at 1mW, while the CO2 laser power varies between 0 to 150mW. We monitor the temperature by Fluorescence Intensity Ratio Thermometry (FIR)[10]. The spectral results are shown in figure 2b. As we increase the thermal load, the temperature rises, and the PL exhibits a blue-shift evolution. This is shown by the reduction in the emission of the low photon-energy peak at 905nm, and enhancement of the high photon-energy peak at 820nm. This trend continues until it reaches the transition temperature of ~1500K. As we increase the temperature further, the emitted photon rate increases sharply at all wavelengths (dotted lines in figure 2b).

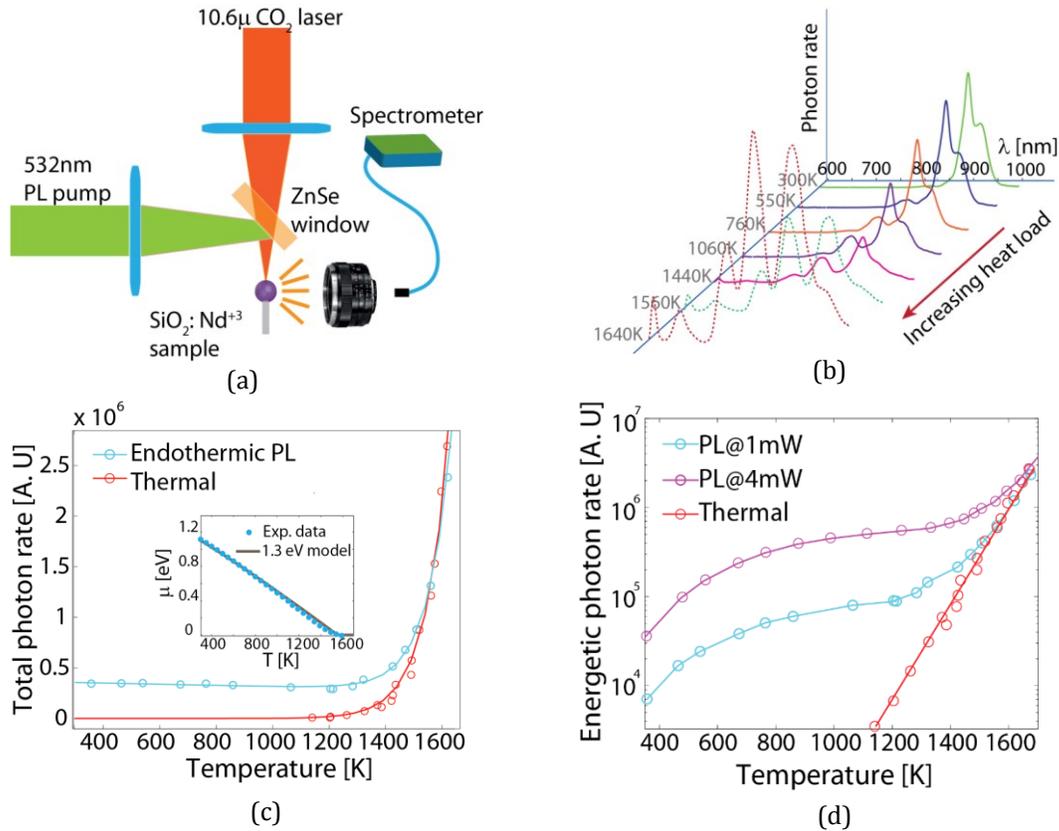

Fig. 2. (a) Experimental setup (b) PL spectra evolution with temperature (c) Total photon rate of PL and thermal emissions (inset: chemical potential vs. temperature) (d) The energetic photon ($\lambda$<850nm) rate for PL and thermal emissions.

Figure 2c shows the total number of emitted photons at wavelengths between 600nm and 1000nm (Blue line). In order to compare PL to thermal emission under equivalent conditions, we turn off the weak PL pump while the thermal current is unchanged (Red line). The thermal emission at temperatures lower than 1150K is below our detection limit and we extrapolate the experimental values down to 300K. The inset of figure 2c shows that the measured chemical potential values (Blue dots) are in good agreement with the theory (Grey line). As demonstrated, the total photon rate is conserved at various temperatures, as long as $\mu > 0$ and increases sharply at $\mu = 0$, while converging with the thermal emission.

The experimental part clearly validates our hypothesis on the nature of PL rate conservation with temperature rise. This can be simply seen by the convergence of the PL and thermal lines in figure 2c: If photon conservation law is not applied, the two separated curves could not converge. Also in figure 2d the order of magnitude enhancement in energetic photons emitted by endothermic PL between 400K and 1100K (Blue and Purple curves) cannot be explained by the contribution of thermal emission (Red curve). Lastly, in agreement with our theoretical model, the reduction of the low-energy PL photon rate (near 905nm), is completely compensated by the increase of PL photons rate at 820nm. Therefore, we demonstrated enhanced spectral blue shift, which is the outcome of energy-balance in a number-conserved photonic system.

**Expending theory of non-thermal radiation to moderate QE:**

Until now we only studied the competition between heat and chemical potential in radiation with high- QE, which conserves the photon flux upon heating. It is critical to understand the thermodynamics under limited QE. In this section we show that although conservation of photon rate doesn't apply, the pump photon rate (or emission rate at 100% QE) is an upper limit for the overall emission rate regardless of the QE. Also, the transition temperature to thermal emission is QE independent. Proving this prediction is important not only to understanding the fundamental thermodynamics of radiation, it will set the upper thermodynamic limit on the efficiency of energy conversion device.

Considering a 2-level system of energy gap $Eg$ and the appropriate transitions, as depicted in Figure 3. $N_1$ includes all vibronic states where the distribution within the $N_1$ band is thermal. In steady state, the detailed balance of incoming and outgoing rates for the $N_1$ band is:

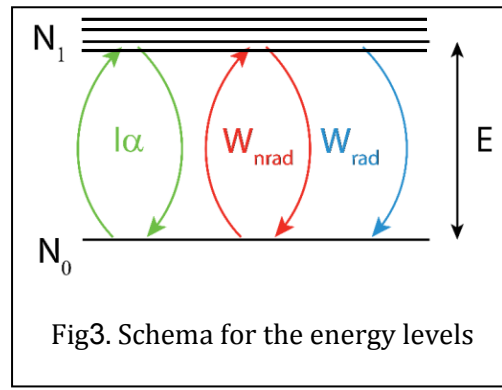

Fig3. Schema for the energy levels

$$N_0 I\alpha + W_{nrad} N_0 e^{\frac{-E_g}{K_b T}} = N_1 (W_{rad} + W_{nrad} + I\alpha)$$

Where: $N_0 + N_1 = N$.

$N_1$ and $N_0$ are the level populations, $I$ is the rate of photons in resonance, $\alpha$ is the absorption coefficient, $W_{rad}$ is the radiative transition rate and $W_{nrad}$ is the non-radiative transition rate, independent on temperature. $N$ is the total number of electrons in the volume. $N_1 I\alpha$ is the stimulated emission rate, which is negligible for PL and thermal radiations. Solving for $N_1$ yields:

$$N_1 = \frac{N\left(I\alpha + W_{nrad} N_0 e^{\frac{-E_g}{K_b T}}\right)}{W_{nrad}\left(1 + e^{\frac{-E_g}{K_b T}}\right) + W_{rad} + I\alpha}.$$

The Quantum efficiency is defined as:
$QE = \frac{W_{rad}}{W_{rad} + W_{nrad}}$, and the PL rate is defined as:
$PL_{rate} = N_1 W_{rad}$. Figure 4 depicts the PL rate versus temperature, for a few $QE$ values: $QE$=1, 0.9, 0.5 and 0.01. As can be seen, for the ideal case, the PL rate is kept constant with the rise in temperature. In this extreme case, $QE$=1, the electron population is essentially isolated from the phonons (giving rise to $W_{nrad} \approx 0$), therefore non-radiative transitions from $N_1$ to $N_0$, **as well as thermal population** of electrons to $N_1$ is prohibited. The PL rate in this case is: $N_1 W_{rad} = NI\alpha W_{rad}/(I\alpha + W_{rad})$, independent on temperature. When the $QE$ is lowered, a relative rate (1-$QE$) is subtracted from this ideal value, as seen by the decrease in the curve's starting values (at cold temperatures). With temperature increase, the PL rate increases, until it intersects the ideal value at T=720 K. At higher temperatures, the emission becomes thermal and joins the thermal-equilibrium emission curve potted in red. Remarkably, all the $QE$ cases intersect the ideal curve at the same point, before the transition to thermal emission. For the extreme low $QE$ case (0.01), the PL curve is very close to the thermal curve, as expected. These simulations show that the critical temperature and the maximum PL rate are $QE$ independent.

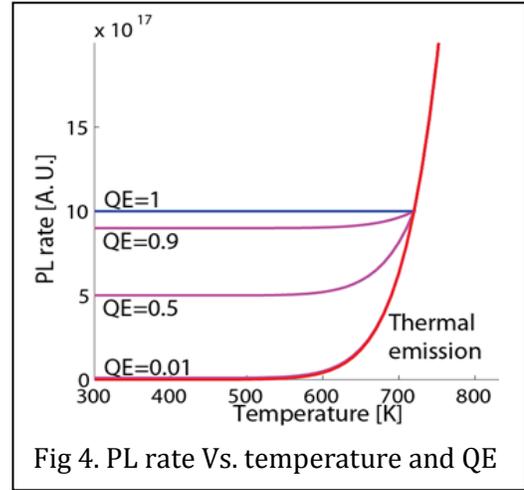

Fig 4. PL rate Vs. temperature and QE

To conclude

We developed the rate equations for PL at moderate $QE$ and found that the maximal PL rate and the transition to thermal emission are independent on the $QE$. In a more general picture, the huge impact of Planck's Black Body formulation is its generality to all thermal emission, of any material, depending only on the temperature (after normalizing the emissivity). As was established above, any PL emission becomes thermal at transition point, where the thermal emission rate equals to the pump rate. That is we found a generalized PL function for any material, where the transition point depends only on the excitation density. These fundamentals of light and heat are expected to have high impact on light and energy science.